\documentclass{article}
\usepackage{amsmath,amssymb}
\usepackage[dvips]{graphicx}

\title{Vlasov simulations of trapping and inhomogeneity in Raman scattering}
\author{D. J. Strozzi$^1$ \thanks{dstrozzi@mit.edu. Present address: LLNL.}
 \and M. M. Shoucri$^2$ \and A. Bers$^1$ \and E. A. Williams$^3$ \and A. B.
 Langdon$^3$ \\
 $^1$Massachusetts Institute of Technology, Cambridge, MA 02139, USA
 \\
 $^2$Institut de Recherche de l'Hydro Qu\'{e}bec, Varennes, Qc J3X1S1,
 Canada \\
 $^3$Lawrence Livermore National Laboratory (LLNL), Livermore, CA 94550,
 USA }   

\date{14 August 2005}

\begin{document}

\maketitle

\begin{abstract}
We study stimulated Raman scattering (SRS) in laser-fusion
conditions with the Eulerian Vlasov code ELVIS. Back SRS from
homogeneous plasmas occurs in sub-picosecond bursts and far exceeds
linear theory. Forward SRS and re-scatter of back SRS are also
observed. The plasma wave frequency downshifts from the linear
dispersion curve, and the electron distribution shows flattening.
This is consistent with trapping and reduces the Landau damping.
There is some acoustic ($\omega\propto k$) activity and possibly
electron acoustic scatter. Kinetic ions do not affect SRS for early
times but suppress it later on. SRS from inhomogeneous plasmas
exhibits a kinetic enhancement for long density scale lengths. More
scattering results when the pump propagates to higher as opposed to
lower density.
 \end{abstract}



\section{Introduction and Code Model}

Laser-plasma interactions must be controlled for inertial fusion to
succeed. This paper examines stimulated Raman scattering (SRS), or
the parametric coupling of a pump light wave (the laser, mode 0) to
a daughter light wave (mode 1) and an electron plasma wave (EPW,
mode 2). Kinetic effects, such as electron trapping, in the daughter
EPW are seen to be important in back SRS (BSRS). 1-D kinetic
simulations presented here show BSRS much greater than coupled-mode
theory for both homogeneous and inhomogeneous plasmas. Strong
nonlinearity and non-thermal electron distributions are seen to
result.

ELVIS \cite{strozzi2004} is a 1-D Eulerian Vlasov code that evolves
the distribution function $f_s$ ($s=$ species; $e$ for electrons) on
a fixed phase-space grid. It uses operator splitting for the time
advance \cite{cheng1976}, \cite{ghizzo1990} and cubic spline
interpolation for shifting $f_s$ in position ($x$) and momentum
($p_x$). Light waves are linearly polarized in $y$. The ions can be
immobile or kinetic. The governing equations are
\begin{equation}
\left[ \partial _{t}+(p_{x}/m_s)\partial _{x}+\left( Z_{s}e\right)
\left( E_{x}+v_{ys}B_{z}\right) \partial _{p_{x}}\right] f_{s} = \nu
_{Ks}\left( x\right) \left( n_{s}\hat{f}_{0s} - f_{s} \right)
\end{equation}%
\begin{equation}
\partial _{x}E_{x}=\frac{e}{\varepsilon _{0}}\sum_{s}Z_{s}n_{s}\qquad \qquad
m_{s}\partial _{t}v_{ys}=eZ_{s}E_{y}
\end{equation}%
\begin{equation}
\left( \partial _{t}\pm c\partial _{x}\right) E^{\pm
}=-\frac{e}{\varepsilon _{0}}\sum_{s}Z_{s}n_{s}v_{ys}\qquad \qquad
E^{\pm }\equiv E_{y}\pm cB_{z}
\end{equation}%
A number-conserving Krook relaxation operator is included, with
relaxation rate $\nu_{Ks}(x)$ and equilibrium Maxwellian
$\hat{f}_{0s}$ ($\int dp\ \hat{f}_{0s}=1)$. We use a large $\nu_{Ks}
\sim 0.2\omega_p$ ($\omega_p^2$ = $n_0e^2/(\epsilon_0 m_e)$) at the
edges of the finite density profile to absorb plasma waves generated
by SRS and prevent reflection. A nonzero central value of $\nu_{Ks}$
can also mimic sideloss from a laser speckle. We advance $E^{\pm }$
without dispersion by shifts of one $x$ gridpoint, which imposes
$dx=c~dt$.

\begin{figure}[t]
 \includegraphics[width=2.3in]{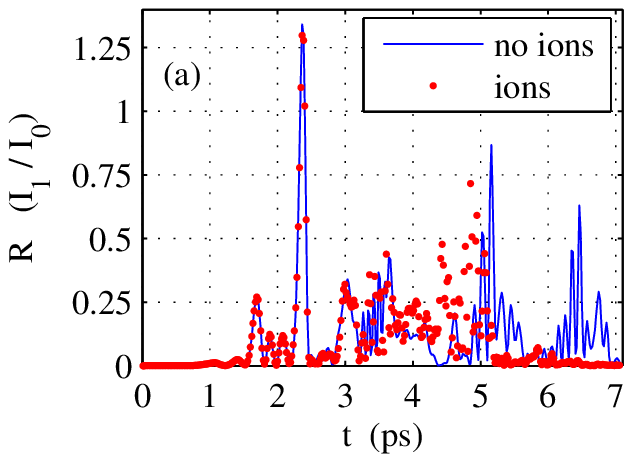}
 \includegraphics[width=2.6in]{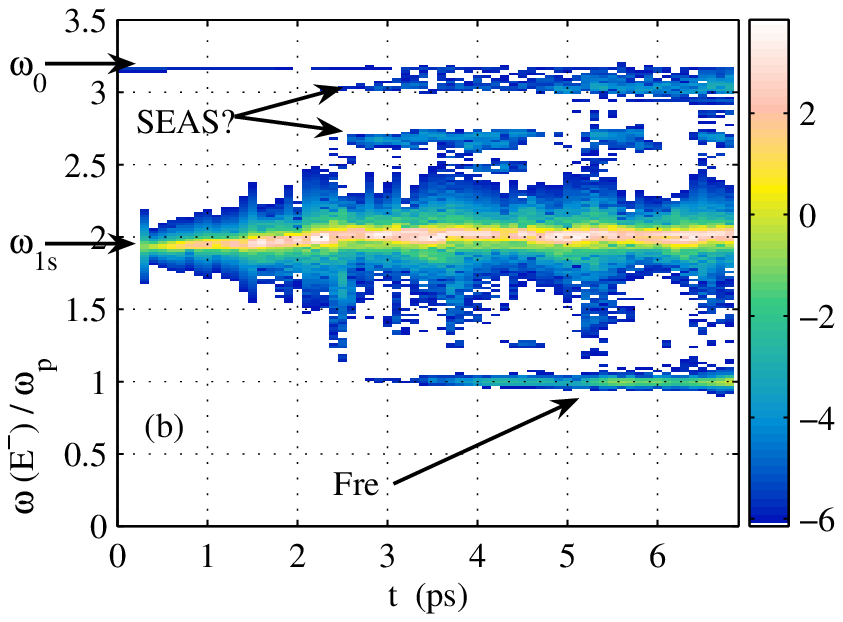}
 \caption{(a) Reflectivity for homogeneous run with immobile (solid curve) and
 kinetic ions (dotted curve) . (b) Spectrum of reflected light for immobile ions.
 ``Fre'' and ``SEAS?'' label BSRS re-scatter and possible electron
 acoustic scattering.}
\end{figure}

\section{Simulations Results}

We simulate a pump laser with $\lambda_0=351$ nm (vacuum) and
intensity $I_0$ = $2\times10^{15}$ W/cm$^2$ impinging from the left
($E^+$ contains the pump) on a finite plasma with a flat central
region 75.1 $\mu$m wide of density $n_0=0.1n_c$ (critical density
$n_c$ = $n_0 \omega_0^2/\omega_p^2$) and temperature $T_e=3$ keV.
Since Vlasov codes are low-noise there are no numerical fluctuations
for SRS to grow from. We therefore inject a counter-propagating seed
light wave via $E^-$ with $\lambda_{1s}=574$ nm and $I_1$ =
$10^{-5}I_0$. This light has the maximum linear BSRS growth rate and
couples to an EPW with $k_2\lambda_D$ = $0.357$ and a Landau damping
rate $\nu_2=0.038\omega_p$ ($\lambda_D$ = $v_{Te}/\omega_p$,
$v_{Te}^2$ = $T_e/m_e$). The $x$ grid spacing is $dx=11.9$ nm, our
algorithm requires $dt=dx/c$, and we use a $p_x$ grid spacing
$dp=0.0437v_{Te}m_e$. SRS is convective for these parameters with a
spatial gain rate $\alpha=0.019$ $\mu$m$^{-1}$, giving a linear
reflectivity $R_{\textrm{\small lin}}$ = $1.72\times10^{-4}$. The
numerical $R$, shown as the solid curve in Fig. 1(a), is well above
this level. $R$ comes in sub-picosecond bursts and has a time
average from 1 ps to the run end of $R_{av}=13.8\%$. $\nu_{Ks}\neq0$
only at the edges. Repeating the run with a nonzero central
$\nu_{Ks}$ shows a sharp cutoff of the reflectivity for
$\nu_{Ks}\gtrsim10^{-3}\omega_p$.

The streaked spectrum of reflected light $E^-$ at the left edge is
shown in Fig. 1(b). Almost all the energy is contained in BSRS.
$\omega_{1s}=1.93\omega_p$ is the seed light frequency. Initially
BSRS occurs at $\omega_{1s}$ but upshifts for $t\gtrsim$ 2 ps,
corresponding to a downshift in $\omega_2$. The weak signal near
$\omega_p$ labeled ``Fre'' is the forward Raman re-scatter of
upshifted BSRS light. The longitudinal $E_x$ spectrum in Fig. 2(a)
reveals the plasma wave at the matching $k$ and $\omega$. Re-scatter
is only possible due to the upshift in $\omega_1$, since $\omega_p$
$>$ $\omega_{1s}/2$. The features slightly below $\omega_0$ and
slightly above $2.5\omega_p$, labeled ``SEAS?'', may be related to
scattering off the acoustic longitudinal activity discussed below
\cite{montgomery2001}. The transmitted light ($E^+$) spectrum (not
presented here) shows for $t\gtrsim$3 ps weak levels of both FSRS
and the anti-Stokes line ($\omega = \omega_0+\omega_p$) of the pump,
even though neither of these is externally seeded.

\begin{figure}[t]
 \includegraphics[width=2.6in]{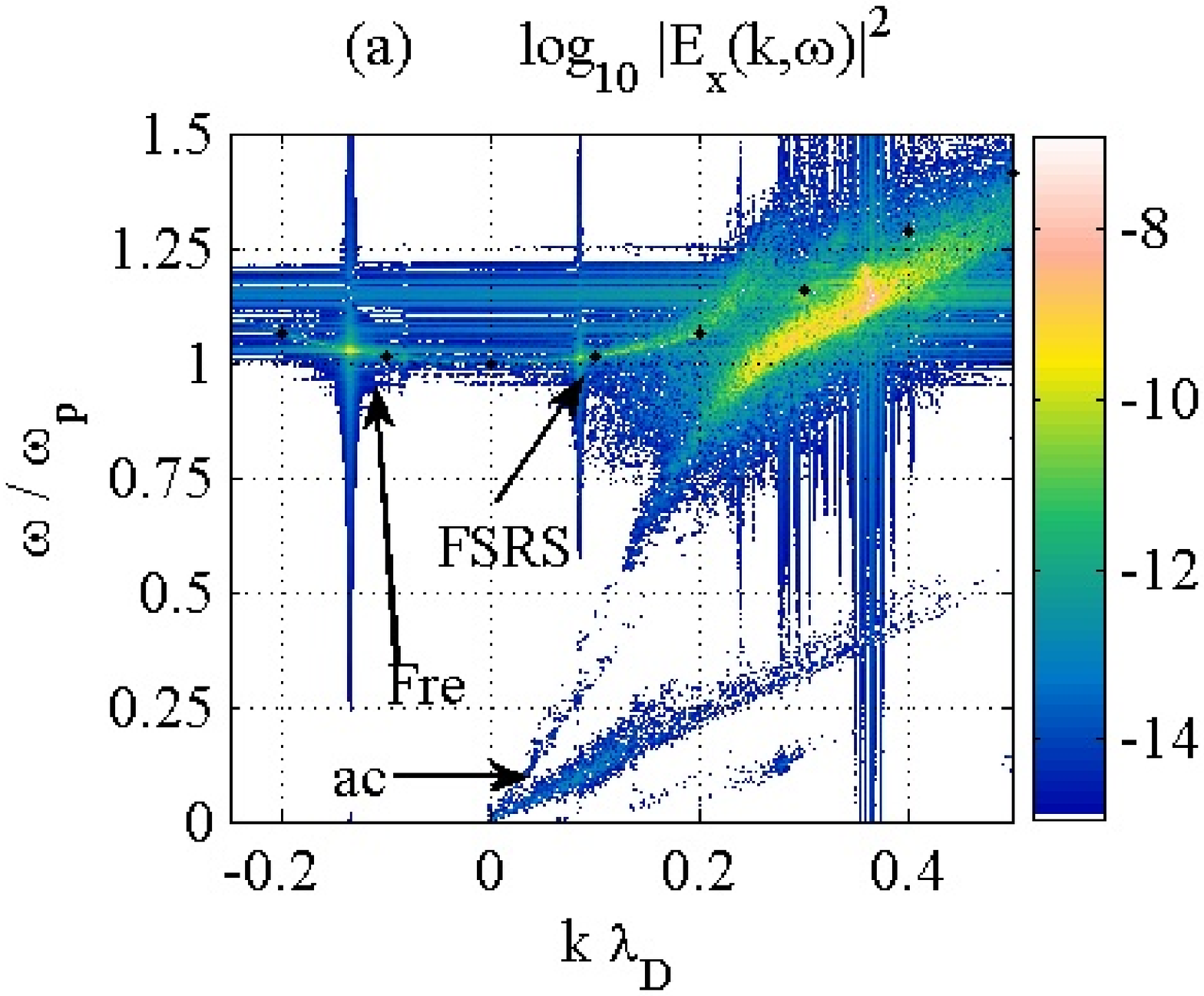}
 \includegraphics[width=2.4in]{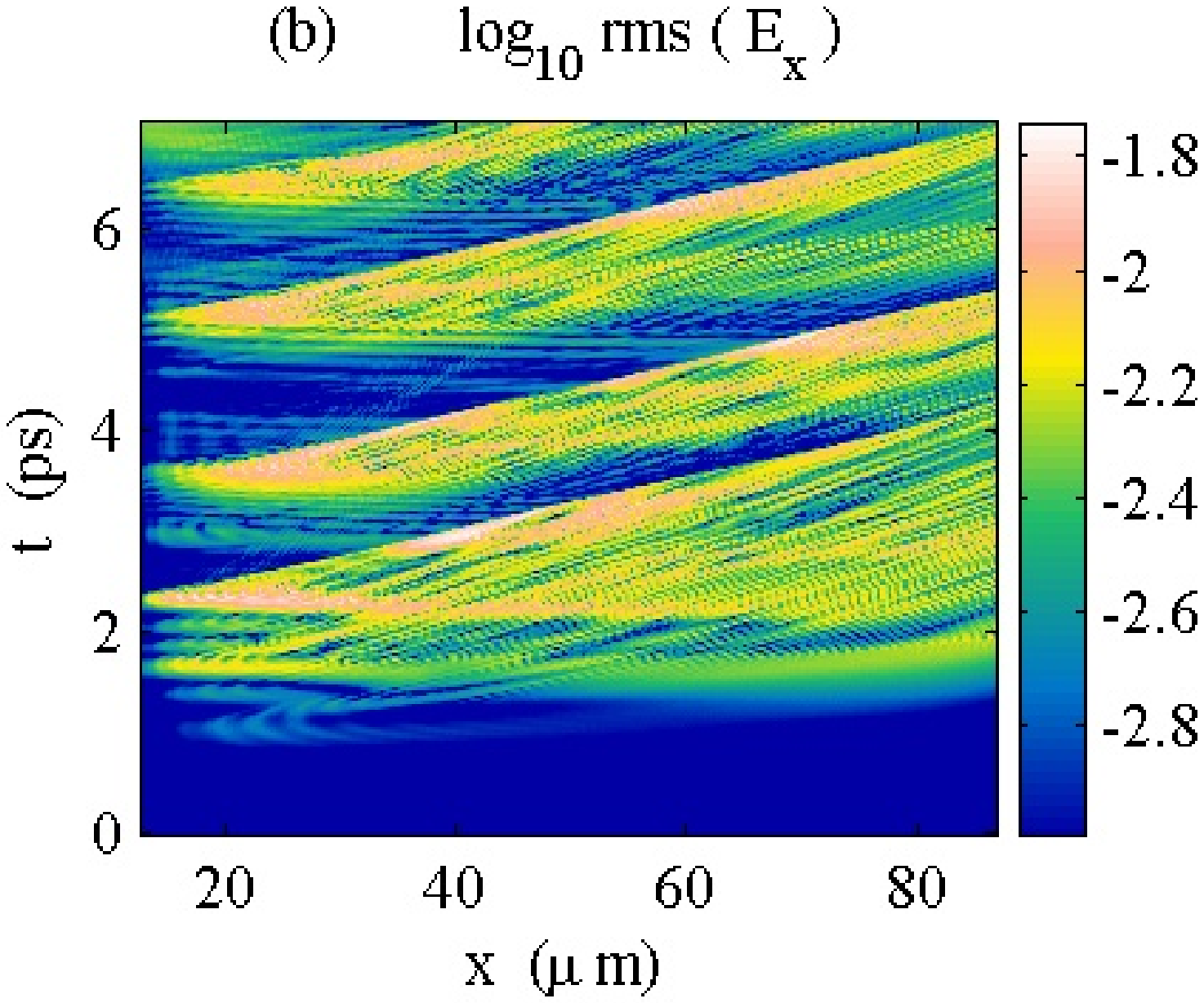}
 \caption{(a) $E_x(k,\omega)$ spectrum for homogeneous, immobile ions run.
 ``Fre'', ``FSRS'', and ``ac'' label BSRS re-scatter, FSRS, and
 acoustic activity. The black dots are the linear EPW dispersion curve.
  (b) rms-averaged $E_x(x,t)$.}
\end{figure}

The longitudinal electric field spectrum $E_x(k,\omega)$, depicted
in Fig. 2(a), reveals that the BSRS plasma-wave activity is
downshifted in frequency from the linear dispersion curve. This is
qualitatively consistent with the frequency downshift due to
electron trapping \cite{morales1972}, and periods of larger
downshift correspond to larger EPW amplitude. The downshifted EPW
connects with an acoustic-like feature ($\omega\propto k$) that
extends to $\omega=0$. There is another, lower phase velocity
acoustic streak, strongest for $\omega \lesssim 0.2\omega_p$. Weak
scattering off them may account for the ``SEAS?'' features in Fig.
1(b). In addition, plasma waves corresponding to FSRS and re-scatter
of BSRS occur on the EPW dispersion curve. Fig. 2(b) presents the
$t$ and $x$ rms-averaged $E_x(x,t)$, which shows the EPWs occur as a
series of wide pulses that move parallel to the laser (that is, to
the right). The group velocity matches the slope of the BSRS plasma
waves. Near the left edge some pulses propagate opposite the laser.

The electron distribution $f_e$ forms phase-space vortices at the
EPW phase velocity $v_{p2}=\omega_2/k_2$ (=0.264$c$ for the linear
EPW). The space-averaged $\langle f_e \rangle$, displayed in Fig.
3(a), is flattened due to trapping in this region. Landau damping
($\sim
\partial f/ \partial v$) is greatly reduced by flattening, which may
thereby enhance the reflectivity \cite{vu2002}. When the EPW
amplitude is large $f_e$ is quite flat, and only for brief periods
($\lesssim0.1$ ps) do we see a small bump (region of
$\partial\langle f_e \rangle /
\partial p_x > 0$) form slightly above $v_{p2}$.

The run was repeated with kinetic helium ions ($m_i=4m_p$, $T_i=750$
eV) and yielded the dotted $R$ in Fig. 1(a). Early in time $R$ is
the same for immobile and kinetic ions, while later in time they
diverge. For the last 2 ps the reflectivity is very low with kinetic
ions. We do not see evidence of Langmuir decay instability (EPW
$\rightarrow$ EPW + IAW (ion acoustic wave)) of the EPW or
stimulated Brillouin scattering (photon $\rightarrow$ photon + IAW)
of the pump. Instead, very high $E_x$ activity develops on the left
edge of the box around $t=5$ ps, involving large ion density
fluctuations; BSRS is minimal after this. Most of the spectral power
in this activity is concentrated in the FSRS and re-scatter of BSRS
features. Further study of the role of ions is underway.

\begin{figure}[t]
 \includegraphics[width=2.4in]{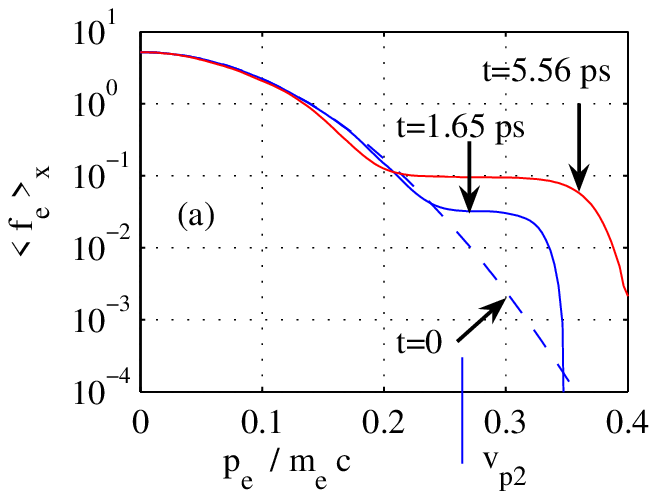}
 \includegraphics[width=2.4in]{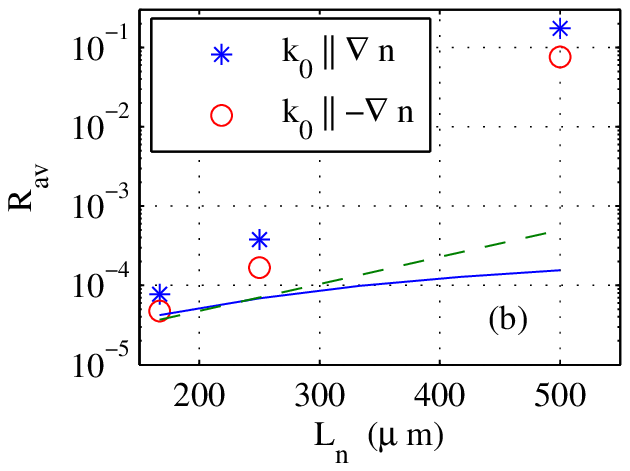}
 \caption{(a) Space-averaged $f_e$ over central 5.62 $\mu$m of box for
 homogeneous, immobile ions run. (b) Average reflectivity for inhomogeneous
 runs with $L_n$ = (167, 250, 500) $\mu$m. (Solid, dashed) curves are the coupled-mode (strong
 damping limit, Rosenbluth undamped) steady-state $R$.}
\end{figure}

In an inhomogeneous medium, the $k$ matching condition $k_0=k_1+k_2$
for a three-wave interaction can only be satisfied at one point.
Away from this point the detuning limits the interaction. We
performed ELVIS simulations for the same parameters as the
homogeneous run with kinetic ions discussed above. However, the
central region of the density profile now has a linear gradient
extending for 100 $\mu$m. We vary the endpoint densities and thereby
change the density scale length $L_n=n/(dn/dx)$. The reflectivities
for several $L_n$ are shown in Fig. 3(b), for the pump propagating
toward higher and lower densities. Also plotted is the steady-state
$R$ predicted from the coupled-mode equations, solved in the limit
of strong damping for the EPW (solid curve) as well as the
Rosenbluth undamped result (dashed curve). Both coupled-mode
calculations give the same $R$ for both directions of pump
propagation, yet the simulation consistently shows higher $R$ for
$\vec{k}_0 || \nabla n$. We are formulating a theory to explain the
high $R$ and the role of pump propagation direction.

\section{Conclusions}

Vlasov simulations of SRS show strong enhancement of the scattering
over coupled-mode predictions for both homogeneous and inhomogeneous
plasmas. The resulting plasma waves do not satisfy the linear
dispersion relation. The electron distribution shows large trapping
and flattening. The role of sideloss and ions need to be further
examined, and analytic models that explain these findings need to be
developed.

This work was supported by the US Dept. of Energy under grant No.
DE-FG02-91ER54109 at MIT and under contract No. W-7405-ENG-48 at
LLNL.


\end{document}